# Selective Etching of Hexagonal Boron Nitride by High-Pressure CF$_4$ Plasma for Individual One-dimensional Ohmic Contacts to Graphene Layers

Yuta Seo[1], Satoru Masubuchi[1,*], Eisuke Watanabe[1], Momoko Onodera[1], Rai Moriya[1], Kenji Watanabe[2], Takashi Taniguchi[1,3], and Tomoki Machida[1,*]

[1] *Institute of Industrial Science, University of Tokyo, 4-6-1 Komaba, Meguro-ku, Tokyo 153-8505, Japan*

[2] *Research Center for Functional Materials,*

*National Institute for Materials Science, 1-1 Namiki, Tsukuba 305-0044, Japan*

[3] *International Center for Materials Nanoarchitectonics,*

*National Institute for Materials Science, 1-1 Namiki, Tsukuba 305-0044, Japan*

*Correspondence: msatoru@iis.u-tokyo.ac.jp, tmachida@iis.u-tokyo.ac.jp

**Abstract:**

We describe a technique for making one-dimensional ohmic contacts to individual graphene layers encapsulated in hexagonal boron nitride (*h*-BN) using $CF_4$ and $O_2$ plasmas. The high etch selectivity of *h*-BN against graphene (> 1000) is achieved by increasing the plasma pressure, which enables etching of *h*-BN, while graphene acts as an etch stop to protect underlying *h*-BN. A low-pressure $O_2$ plasma anisotropically etches graphene in the vertical direction, which exposes graphene edges at *h*-BN sidewalls. Despite the $O_2$ plasma bombardment, the lower *h*-BN layer functions as an insulating layer. Thus, this method allows us to pattern metal electrodes on *h*-BN over a second graphene layer. Subsequent electron-beam lithography and evaporation fabricate metal contacts at the graphene edges that are active down to cryogenic temperatures. This fabrication method is demonstrated by the preparation of a graphene Hall bar with a graphite back-gate and double bilayer-graphene Hall bar devices. The high flexibility of the device geometries enabled by this method creates access to a variety of experiments on electrostatically coupled graphene layers.

Recent developments in the assembly of van der Waals (vdW) heterostructures[1], techniques based on the sequential pick-up and drop-off release of atomic layers[2], have enabled the fabrication of complex vdW heterostructures comprising multiple channel layers[3,4]. The preparation of graphene/*h*-BN heterostructures consisting of multiple graphene layers, to obtain high-mobility field-effect transistors[5] and tunneling devices[6], is now a standardized process. In conventional semiconductor-based two-dimensional multiple quantum wells[7], Coulomb interactions between electrons result in the emergence of electronic ground states such as exciton-condensate[8] and quantum Hall ferromagnetic[9] states. Combined with the growing opportunity to fabricate complex vdW heterostructures, there is considerable demand for the development of fabrication techniques for individual metal contacts for each graphene layer in vdW heterostructures.

The interfaces between metals and two-dimensional materials can be classified into two categories according to their geometrical configurations: two-dimensional top-contact interfaces and one-dimensional edge-contact interfaces[10]. One-dimensional contact geometry provides higher flexibility of device geometries than two-dimensional contacts, because of the reduced dimension in the interface between metal and graphene channels. One-dimensional edge-contact fabrication techniques have been developed for graphene[2]. In general, the fabrication of edge contacts occurs via two steps: (i) the dry etching of heterostructures to expose the encapsulated graphene edges in hexagonal boron nitride (*h*-BN), and (ii) the subsequent deposition of metal electrodes by evaporation at the exposed edges. However, the intrinsically thin nature of exfoliated *h*-BN flakes and high *h*-BN etching rate for conventional plasma source ($CHF_3$) requires precise control of the etching time in a unit of seconds[11] to achieve the depth control. This requirement is often avoided by offsetting the graphene layers in the lateral directions so that the contacts can be formed at the different positions[12].

The recent advent of complicated vdW heterostructures consisting of multiple channel regions[3] has created the demands for addressing individual layers spaced by thin (~10 nm) *h*-BN dielectric layers. Recent studies report that the graphene exhibits significant resistance against some fluorinating plasma sources, which provides an opportunity to selectively etch graphene and *h*-BN while utilizing graphene layer as an etch stop to protect underlying *h*-BN[13-15]. Exposure to $XeF_2$ and subsequent metal evaporation have been demonstrated to fabricate ohmic contacts for graphene channels[13]. However, these techniques[13,14] rely on the interface between metal electrodes and fluorinated graphene, which requires two-dimensional contact area. In addition, transport takes place through fluorinated graphene, and under these conditions, precise band-alignment modeling for the metal and channel layers is difficult. The gas source previously utilized to demonstrate fabrication ($XeF_2$) is not commonly available in most user-facility cleanrooms. The use of $SF_6$ has also been demonstrated to implement the etch-stop recipe[15]. The conventional plasma source $CF_4$ can be used for selective etching of graphene and *h*-BN[14], however the achieved selectivity ~100 (for *h*-BN and bilayer graphene) is not sufficient for etching graphene/*h*-BN heterostructure which typically comprises thin 0.3 nm graphene and thick 10-40 nm *h*-BN, which requires etching selectivity higher than 10-40/0.3 = 30-130. In addition, the insulation of underlying *h*-BN region below the graphene layer exposed to the plasma sources has been unclear, and the patterning of the metal electrodes over exposed area has not yet been demonstrated.

In this work, we establish a method to realize selective etching of *h*-BN by commonly available plasma source $CF_4$. The etching rates of graphene (*h*-BN) by $CF_4$ plasma decreased (increased) with the plasma pressure ($P$). At high pressure condition, $P = 10$ Pa, the etch selectivity of *h*-BN against graphene reaches >3000. A subsequent low-pressure $O_2$ plasma anisotropically etches graphene in the vertical direction, which exposes graphene edges at *h*-BN sidewalls. Despite the $CF_4$ plasma and subsequent $O_2$ plasma bombardment, the lower *h*-

BN layer remains insulating. The technique allows us to fabricate one-dimensional contacts, addressing each graphene layer without using two-dimensional contacts.

Fig. 1(a)-(h) schematically illustrate the fabrication process of a $h$-BN/graphene/$h$-BN/graphite sample using $CF_4$ and $O_2$ plasmas. Fig. 1(i)-(l) are optical microscopy (OM) images of the sample acquired during the fabrication process. The heterostructures are prepared using the robotic assembly system for vdW heterostructures [3,16] [Fig. 1(a)]. The channel region is patterned using a hydrogen silsesquioxane/poly(methyl methacrylate) (HSQ/PMMA) etching mask and high-pressure ($P$ = 10 Pa) $CF_4$-based plasma [Fig. 1(b)-(d)]. Since $CF_4$ selectively etches $h$-BN, the lower $h$-BN layer, covered by graphene, is not etched. Thus, $h$-BN terraces are formed in the regions covered by graphene [Fig. 1(e)]. Next, a low-pressure $O_2$ plasma etches the graphene on the $h$-BN layer [black arrow in Fig. 1(e)]. In this case, the etching takes place vertically, and thus graphene edges are exposed at the intersection of the upper and lower $h$-BN layers [black arrow in Fig. 1(g)]. Finally, metal evaporation—first Cr (5, 5, and 2 nm), followed by Au (10, 10, and 30 nm), both layers at +30°, –30°, and 0°—generates one-dimensional ohmic contacts for the graphene layer. Note that the metal electrodes are patterned on the $h$-BN above the graphite backgate [Highlighted by red lines in Fig. 1(h)]. In contrast to the previously reported etching techniques, this technique obviates the need for the lateral offsetting between the metal electrodes and graphite backgate.

The key parameter for the effective implementation of the process we present is pressure of the plasma chamber ($P$). In Fig. 2, we show the etch rates of (a) graphene and (b) $h$-BN as a function of $P$ (The methods to extract the values are presented in the supplementary information). When the pressure was increased from 0.7 Pa to 10 Pa, the etch rate of graphene was decreased from 3.5 pm/s to 0.9 pm/s, while that of $h$-BN was increased from 0.75 nm/s to 3.0 nm/s. The selectivity between graphene and $h$-BN, $S$ = (the etch rate of $h$-BN)/(the etch rate of graphene), was increased from 230 to 3500 [Fig. 2(c)]. The enhancement of the etching selectivity by an order provides opportunity for fabricating the proposed device geometry.

In Fig. 2 and Fig. S3 in supplementary information, we compare the responses of the *h*-BN/graphene/*h*-BN/graphite heterostructures for differing pressure (0.7-10 Pa) plasma comprising $CF_4$ and $O_2$. Fig. 2(d) and (e) show OM images of *h*-BN/graphene/*h*-BN/graphite before etching at 10 and 0.7 Pa. These samples are etched in the inductively coupled plasma - reactive ion etching (ICP-RIE) system with 20-sccm $CF_4$ and 2-sccm $O_2$ for 180 s, using with 30 W of ICP power and 0 W of bias power. Note that no etching masks are utilized here. After plasma etching, the regions of the lower layer of *h*-BN that were not covered by graphene (indicated by dashed lines) are completely etched for $P = 0.7$-10 Pa [Fig. 2(f), (g), and Fig. S3(e)-(h).]. The parts of the lower *h*-BN layer covered by graphene exhibit different textures. At high pressures ($P = 4$-10 Pa), the surface is smooth, while at low pressure ($P = 0.7$ Pa), the graphene surface is distorted and more transparent than that at $P = 4$-10 Pa [the regions highlighted by white dashed lines in Fig. 2(f) and (g)].

To observe the microscopic features of the remaining lower *h*-BN terrace, we acquired scanning electron microscopy (SEM) images of the *h*-BN terraces formed after $CF_4/O_2$ etching. The SEM measurements were performed at the positions indicated by the white arrows in the insets of Fig. 2(h) and (i). For the sample treated at $P = 10$ Pa, the surface of *h*-BN terrace structure is flat, and no discernible holes are observed down to the resolution limit of the SEM system (~10 nm). The surface smoothness was also confirmed by AFM measurements, which shows root-mean-square surface roughness of ~84 pm [Fig. S4(a) in the supplementary information], which is comparable to that of bare *h*-BN surface ~80 pm[17]. The Raman spectrum taken after etching exhibits prominent G peaks [Fig. S6(a) in the supplementary information], indicating that the etching preserves graphene's hexagonal lattice. Simultaneously, the D peaks emerged, and the 2D peaks were diminished, indicating the fluorination of graphene flake.

Therefore, the fluorinated graphene exhibits significant resistance against $CF_4$ plasma, and protects the underlying *h*-BN. This trend is preserved for lowering pressure down to 4 Pa [Figs. S3-6 in the supplementary information]. In contrast, for the sample treated at $P$ = 0.7 Pa, the terrace structure is composed of many island-shaped *h*-BN structures, and a large number of holes are formed in the bottom *h*-BN layer. The G peaks were diminished [Fig. S6(d) in the supplementary information], which indicates deterioration of graphene at $P$ = 0.7 Pa. This result indicates that the pressure can be utilized as a tuning knob to control the etch resistance of graphene.

We provide a possible explanation for the pressure-dependent etching behavior by considering the average energy and mean free path of positive ions in the plasma. In general, plasma etching is caused by two mechanisms (i) chemical reactions and (ii) physical sputtering via the bombardment of positive ions that are accelerated by the electric field that develops on the surface[18]. At high pressure, the accelerated positive ions frequently collide with other ions, and thus the mean free path and average momentum are smaller than those at low pressure [Fig. 2(j) and (k)]. The smaller mean free path and momentum reduce sputtering rates, and etching proceeds predominantly via chemical reactions[19]. The etching of *h*-BN takes place through chemical reaction with fluorine and reactant is in a gas form. On the other hand fluorinated graphene is stable under ambient conditions[20], and the chemical etch rate of graphene becomes smaller than that of *h*-BN [Fig. 2(j)]. With increasing the pressure, the probability of physical sputtering rate decreases[19]. Thus, graphene exhibits better resistance against fluorinating plasma. The formation of island-shaped *h*-BN residues [Fig. 2(i)] at low pressure can be explained by considering the defect formation through the physical sputtering of graphene and subsequent chemical etching of underlying *h*-BN. Since the plasma etching behavior is common to the ICP-RIE system, it should be possible to reproduce our results in other etching systems by adjusting the power according to the chamber sizes.

By using the etching process described above, we demonstrate the fabrication of three-types of devices (i) a $h$-BN/graphene/$h$-BN/graphite bar-shaped device, (ii) a $h$-BN/bilayer graphene/$h$-BN/graphite Hall-bar device, and (iii) a $h$-BN/bilayer graphene/$h$-BN/bilayer graphene/$h$-BN double Hall-bar device. In all these devices, one-dimensional contacts were patterned over lower graphite/graphene layers. First, we show an OM image of the bar-shaped device [Fig. 3(a)-(b)]. Fig. 3(c)-(e) show the two-terminal resistance $R_{2t}$ as a function of the back-gate bias voltage $V_{bg}$ for various channel lengths, $L$ = (c) 3 μm, (d) 5 μm, and (e) 7 μm. Note the small $R_{2t} <$ 1.5 kΩ at the charge carrier density $n = \pm 1.9 \times 10^{12}$ cm$^{-2}$, which indicates that the contact resistance is smaller than 750 Ω, which is sufficient for conducting measurements of quantum Hall effects using a two-terminal geometry [Fig. S10 in the supplementary information]. Fig. S11(b) in the supplementary information shows current-voltage characteristics measured between contacts 1 and 2 [Fig. S11(a) in the supplementary information]. The $I$-$V$ characteristics are linear up to 0.1 V at 1.7 K, which confirms the ohmic contacts. Fig. S12 shows the four-terminal quantum Hall measurements of $h$-BN/bilayer graphene/$h$-BN/Graphite sample. The observation of clear spin and valley splitting quantum Hall effects confirms the conduction of quantum transport measurements, which demonstrates the usability of the one-dimensional contact for characterizing quantum-transport properties.

From the measured $R_{2t}$ we discuss the distribution of contact resistance $R_c$. The estimation of $R_c$ using the transfer length method (TLM) proved to be challenging because our $h$-BN/graphene/$h$-BN device exhibits high charge carrier mobility (~100,000 cm$^2$/Vs) and hence the transport is in the ballistic regime. The ballistic transport was confirmed by the magnetic focusing effect measurements in a Hall-bar geometry sample [Fig. S13 in the supplementary information]. This means that the two-terminal resistance does not scale with channel length $L$ and width $W$ according to $R_{2t} = \rho(L/W)$, where $\rho$ is the channel resistivity [Fig. 3(f)]. Therefore, we employed a simpler method, which gives an upper-bound estimate of the contact resistance $\widehat{R_\mathrm{c}} = R_\mathrm{2t}/2$ for each device, unlike TLM. Because of the high mobility, we can assume that the contribution of $\rho$ to the two-terminal resistance $R_\mathrm{2t}$ becomes negligible, compared to the contact resistance, at high $n$. From the $\widehat{R_\mathrm{c}}W$ values extracted from seven devices, we estimate a median $\widehat{R_\mathrm{c}}W$ of 500 Ω μm at $n = 1.9 \times 10^{12}$ cm$^{-2}$ [Fig. S9 in the supplementary information]. Considering that the values of $R_\mathrm{c}$ obtained in the previous one-dimensional edge contacting methods[2] were $R_\mathrm{c} \sim 150$ Ω μm, and our estimate provides an upper bound for $R_\mathrm{c}$, we consider that our method achieved sufficiently transparent ohmic contacts that can be utilized to investigate fundamental quantum-transport properties of graphene-based devices.

Next, we show the $h$-BN/bilayer-graphene/$h$-BN device with a graphite backgate [inset of Fig. 4(a)]. The four-terminal longitudinal resistance ($R_{xx}$) exhibits ambipolar field effects up to $V_\mathrm{bg} = \pm 14$ V [Fig. 4 (a)]. From the conductivity $\sigma$ vs. $n$ curve, we estimate the residual charge carrier inhomogeneity $n^*$ to be $3 \times 10^{10}$ cm$^{-2}$ [Fig. 4(b)]. The charge carrier mobility $\mu$ reaches 100,000 cm$^2$/Vs [Fig. 4(c)]. Note that during the measurements no appreciable leak current ($I_\mathrm{leak}$) was observed up to $V_\mathrm{bg} = \pm 11$ V [Fig. 4 (d)], and $I_\mathrm{leak}$ starts to increase at $V_\mathrm{bg} > 13$ V and $V_\mathrm{bg} < -14$ V. The observed high breakdown voltage allows us to inject carriers up to $n = \pm 1 \times 10^{13}$ cm$^{-2}$, which can be exploited to investigate quantum transport properties of graphene. In general, the breakdown voltage of a pristine $h$-BN flake is 1.2 V/nm[21]. In our device, the thickness of the $h$-BN is 31 nm; thus, this observation indicates that our lower $h$-BN layer remains insulating up to 0.45 V/nm, even after oxygen plasma bombardment in the final etching step [Fig. 1(f)].

Finally, by repeating the etching process depicted in Fig. 1(b)-(g), we fabricated a double Hall-bar device containing bilayer graphene [Fig. S17(a) in the supplementary information] with one-dimensional contacts for both the top and bottom graphene bilayers [Fig. S17(b) in the supplementary information]. The individual contacts for the graphene layers allow us to apply the top-gate and back-gate bias voltages ($V_{tg}$ and $V_{bg}$) to the bilayer graphene [Fig. S17(c) in the supplementary information]. In the color plot of longitudinal resistance $R_{xx}$ as a function of $V_{tg}$ and $V_{bg}$, the resistance peak spans a diagonal. The resistance peak becomes more intense as $V_{tg}$ deviates from 0 V. [Fig. S17(d) in the supplementary information]. These results indicate the control of the charge carrier density by $V_{tg}$ and $V_{bg}$ and the development of a vertical electric field, thereby confirming the electrical insulation between graphene layers. These observations demonstrate the ability to make contacts for individual graphene layers embedded in $h$-BN, without disrupting the insulating behavior of $h$-BN.

In this work we presented a method to make individual one-dimensional contacts for graphene layers encapsulated in $h$-BN. The sequential etching by high-pressure $CF_4$ and low-pressure $O_2$ allows us to selectively etch $h$-BN and fabricate one-dimensional exposed graphene edges, respectively. The ability to access each individual graphene layer via a one-dimensional contact provides access to a variety of experiments on complex graphene/$h$-BN vdW heterostructures.

*Data Availability*

The data that support the findings of this study are available from the corresponding author upon reasonable request.

*Supplementary Material*

See supplementary material for details on methods used. Supplementary figures showing the etching in progress, two-terminal conductance plots, and details of double Hall-bar device containing bilayer-graphene are also provided, and supplementary data from the transport measurements is given in the supplementary table.

*Acknowledgements*

This work was supported by the Core Research for Evolutional Science and Technology, Japan Science and Technology Agency (JST) under Grant No. JPMJCR15F3, and by the Japan Society for the Promotion of Science (JSPS) KAKENHI under Grant Nos. JP19H01820, JP19H02542, JP20H00127, and JP20H00354.

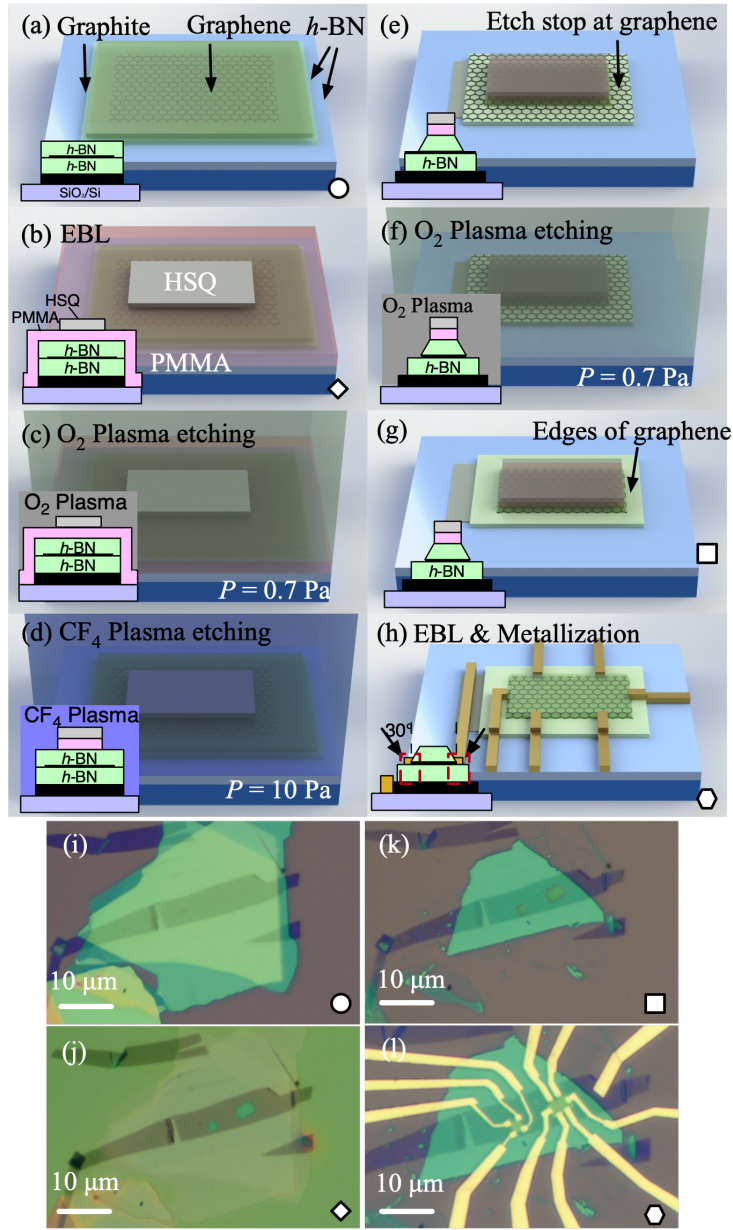

Fig. 1. *h*-BN/graphene/*h*-BN/graphite heterostructure device fabrication. (a) Schematic illustration of the *h*-BN/graphene/*h*-BN/graphite heterostructure. (b) PMMA and HSQ are spun at 4000 rpm. HSQ is patterned by electron beam lithography (EBL). (c) The sample is then loaded into an inductively coupled plasma (ICP) reactive ion etching (RIE) system (SAMOCO, RIE-200iP). The PMMA layer is etched in an $O_2$ plasma with flow rate of 30 sccm at pressure $P = 0.7$ Pa with HSQ as the etch mask. The etch rate of PMMA is approximately 16 nm/min with 33 W of ICP power and 0 W of bias power. (d) The heterostructure is etched by $CF_4$ and $O_2$ plasmas with flow rates of 20 sccm and 2 sccm, respectively, at pressure $P = 10$ Pa. (e) The etching does not proceed through graphene. (f) O$_2$ plasma ($P$ = 0.7 Pa, 33 W ICP power, 0 W bias power) removes graphene. (h) PMMA layer is dissolved by immersing the sample in acetone thereby removing HSQ. The second electron-beam lithography defines metal electrode patterns. Cr (5 + 5 + 2 nm) is deposited followed by Au (10 + 10 + 30 nm); both layers at +30°, –30°, and 0°, respectively, to form 1D edge contacts. (i)-(l) Optical microscopy image of the heterostructure at the fabrication stages shown in (a), (b), (g), and (h), respectively.

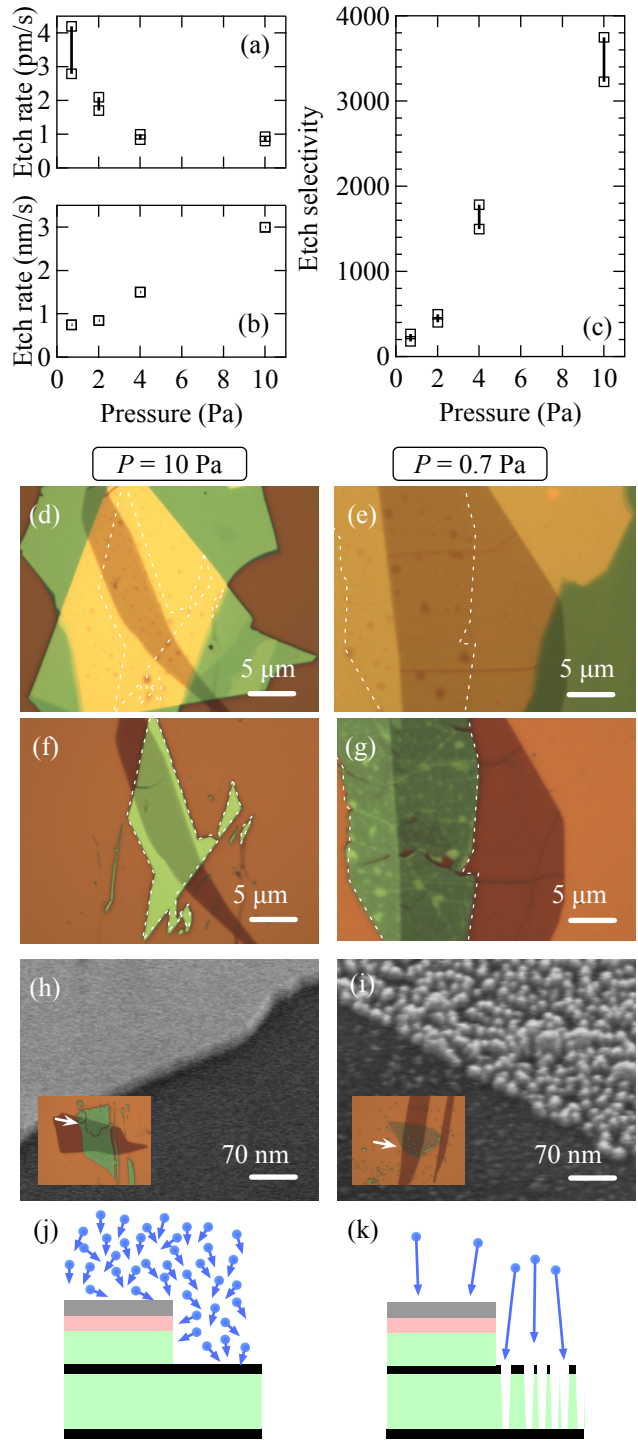

Fig. 2. (a), (b) The etching rates of (a) graphene and (b) *h*-BN as a function of pressure. (c) The etch selectivity as a function of pressure. (d)-(g) OM images of *h*-BN/graphene/*h*-BN before [(d), (e)] and after [(f), (g)] etching at [(d), (f)] *P* = 10 Pa and [(e), (g)] *P* = 0.7 Pa. The outlines of graphene flakes are highlighted by white dashed lines. (h) and (i) SEM images of *h*-BN/graphene/*h*-BN after plasma etching at *P* = 10 Pa (h) and 0.7 Pa (i). The etching time is 150 s. The images depict the boundary regions of the graphene flakes indicated by white arrows in the larger OM views shown as insets. (j) and (k) Schematic illustrations of *h*-BN/graphene/*h*-BN heterostructures in (j) high- and (k) low-pressure plasma, where their mean free path (and energy) are small and large, respectively.

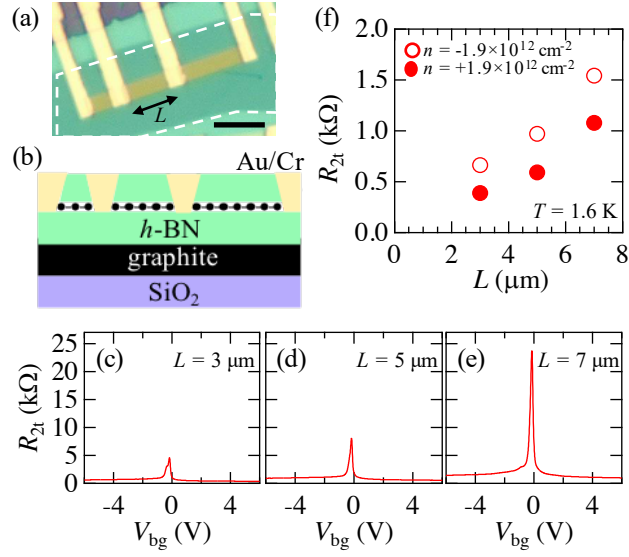

Fig. 3. (a) Optical microscopy image and (b) schematic of the device with graphite backgate. The scale bar corresponds to 5 μm. (c)-(e) $R_{2t}$ as a function of $V_{bg}$ for devices with channel length (c) $L = 3$ μm, (d) 5 μm, and (e) 7 μm. (f) $R_{2t}$ at $n =$ (filled circle) $1.9 \times 10^{12}$ cm$^{-2}$ and (open circle) $-1.9 \times 10^{12}$ cm$^{-2}$ as a function of $L$.

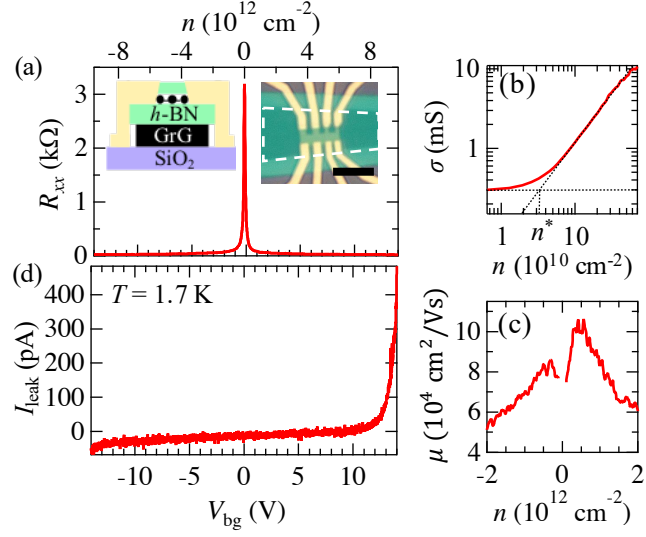

Fig. 4. *h*-BN/bilayer-graphene/*h*-BN device with a graphite backgate (GrG). (a) $R_{xx}$ as a function of $V_{bg}$ and $n$. The left and right insets show a schematic and an optical microscopy image (the scale bar is 5 μm), respectively, of the device. The graphite backgate is highlighted by white dashed lines. (b) $\sigma$ as a function of $n$. (c) $\mu$ as a function of $n$. All measurements were performed at $T = 1.7$ K. (d) $I_{leak}$ as a function of $V_{bg}$.